\documentclass[twocolumn, showpacs, amsmath,amssymb, pra]{revtex4}

\usepackage{graphicx}
\usepackage{times}

\newcommand{\bra}[1]{\langle #1|}
\newcommand{\ket}[1]{| #1 \rangle}
\newcommand{\op}[1]{\hat{#1}}

\newcommand{\cre}[1]{\hat{#1}^\dagger}
\newcommand{\des}[1]{\hat{#1}}
\newcommand{\expi}[1]{ {\rm e}^{{\rm i} #1}}

\begin{document}

\title{Superposition states of ultracold bosons in rotating rings with a realistic potential barrier}

\author{Andreas Nunnenkamp}
\affiliation{Departments of Physics and Applied Physics, Yale University, New Haven, CT 06520, USA}
\author{Ana Maria Rey}
\affiliation{JILA, NIST and Department of Physics, University of Colorado, Boulder, CO 80309, USA}
\author{Keith Burnett}
\affiliation{University of Sheffield, Firth Court, Western Bank, Sheffield, S10 2TN, UK}

\date{\today}

\begin{abstract}

In a recent paper [Phys.~Rev.~A \textbf{82}, 063623 (2010)] Hallwood \textit{et al.}~argued that it is feasible to create large superposition states with strongly interacting bosons in rotating rings. Here we investigate in detail how the superposition states in rotating-ring lattices depend on interaction strength and barrier height. With respect to the latter we find a trade-off between large energy gaps and high cat quality. Most importantly, we go beyond the $\delta$-function approximation for the barrier potential and show that the energy gap decreases exponentially with the number of particles for weak barrier potentials of finite width. These are crucial issues in the design of experiments to realize superposition states.
\end{abstract}

\pacs{67.85.Hj; 03.75.Gg; 03.75.Lm}

\maketitle

Ultracold bosons in rotating ring traps are a fascinating subject of research because the particles in this system can form various multi-particle superposition states. Some of these superposition states may play a central role in studies of the quantum-to-classical transition~\cite{Leggett2002} or in applications such as entanglement-enhanced metrology~\cite{Leibfried2004, Giovannetti2006, Pezze2009}.

Most of the work in this area has focused on the production of so-called NOON states in uniform ring lattices~\cite{Hallwood06a,Rey07} and superlattices~\cite{Nunnenkamp08}. In these systems there are critical rotation frequencies at which the single-particle spectrum is degenerate so that even weak interactions lead to strong correlations between the particles. The energy gap between ground and excited states for these systems does, as one might expect for multi-particle transitions, decrease exponentially with the number of particles. This does, of course, limit the schemes proposed to generate these quasi-momentum NOON states to operate with a modest number of atoms~\cite{Rey07,Hallwood06b}.

Very recently, the authors of Ref.~\cite{Hallwood2010} showed that for strong interactions the ground state of the Lieb-Liniger model with a $\delta$-function potential barrier is a superposition of states with different total (angular) momenta of the particles. They found that in this regime the energy gap is independent of the number of particles for small barrier heights and proportional to the number of particles for strong barrier heights. They come to the conclusion that the production of superposition states with much larger number of particles will become feasible.

In this paper we study in detail the different superposition states that occur for ultracold bosons in rotating ring lattices with a realistic potential barrier. We show that these superposition states strongly differ in the regimes of weak and strong interaction strength (see also Ref.~\cite{Hallwood2010}) as well as of weak and large potential barriers. Most importantly, we demonstrate that for weak barriers the energy gap decreases exponentially with the number of particles as soon as the finite width of the potential is taken into account. These are important limitations to producing superposition states that must be taken into account in planning experiments. 

We focus on systems at a critical rotation frequency where the single-particle spectrum of the uniform ring is doubly degenerate and introduce a potential barrier that breaks translational invariance leading to superposition states of different quasi-momenta. We study systems with much less than one atom per site in order to avoid the superfluid-to-Mott-insulator transition for strong interactions. At low filling factors the particles only feel the quadratic part of the lattice dispersion so that our lattice simulations should be relevant for a continuous loop as studied in Ref.~\cite{Hallwood2010}.

We use both exact diagonalization of systems with small numbers of atoms and sites and the Bose-Fermi mapping in the Tonks-Giradeau regime to characterize the many-body ground state. Going beyond Ref.~\cite{Hallwood2010}, we calculate the ground-state overlap with three different trail wavefunctions, the spectrum of the single-particle density matrix, and the experimentally accessible correlation functions, such as momentum distribution and momentum noise correlations.

Like Ref.~\cite{Hallwood2010} we find that the superposition states strongly depend on the interaction strength. For weak potential barriers and weak interactions it is a Bose-Einstein condensate, for intermediate strength it is a strongly correlated NOON state, and for strong interactions, where the particles are locked together, it is the superposition of two center-of-mass motional states. In the lattice case we consider that the superpositions also depend on the potential height. We find that they are degraded significantly if the potential barrier is large, which leads to a trade-off between the energy gap and cat quality.

The limit of strong interactions and weak potential barrier is most attractive for generating superpositions with many particles. In this case, the energy gap is independent of the number of particles for a single-site barrier~\cite{Hallwood2010}. However, as soon as we consider a physical potential with finite width, the energy gap decreases exponentially with increasing number of particles. Since any real potential will have a finite width this will give a physical limit to the number of particles in the superposition state.

\emph{Hamiltonian.} We consider a system of $N$ ultracold bosons with mass $M$ confined in a 1D ring lattice of $L$ sites with lattice constant $d$. The ring is rotated in its plane with angular velocity $\Omega$. In the rotating frame the Hamiltonian is~\cite{Bhat2006, Rey07}
\begin{equation}
\op{H}
= \sum_{j=1}^L \left[ -J \left(\expi{\theta} \cre{a}_{j+1} \des{a}_j + \mathrm{H.c.} \right) + \frac{U}{2} \hat{n}_j(\hat{n}_j-1) + V_j \hat{n}_j \right]
\label{Ham}
\end{equation}
where $\hat{n}_j=\hat{a}_j^{\dagger}\hat{a}_{j}$ and $\hat{a}_{j}$ are the number and bosonic annihilation operators of a particle at site $j$, $\theta = M \Omega L d^2/h$ is the phase twist induced by rotation, $J$ is the hopping energy between nearest-neighbor sites, $U$ the on-site interaction energy, and $V_j$ describes the potential barrier at site $j$.

To understand the effect of rotation on the atoms in the ring lattice, we write the many-body Hamiltonian (\ref{Ham}) in terms of the quasi-momentum operators $\hat{b}_q = \frac{1}{\sqrt{L}} \sum_{j=1}^{L} \hat{a}_j e^{-2\pi i q j/L}$, where $2\pi q / d L$ is the quasi-momentum and $q=0,\dots,L-1$ an integer. In this basis the Hamiltonian (\ref{Ham}) has the form~\citep{Hallwood06a, Rey07}
\begin{align}
\hat{H} =
&\sum_{q=0}^{L-1} E_q\hat{b}_q^{\dagger}\hat{b}_{q}+\frac{U}{2L}
\sum_{q,s,l=0}^{L-1}\hat{b}_q^{\dagger} \hat{b}_s^{\dagger}
\hat{b}_l\hat{b}_{[q+s-l] \, \textrm{mod} \, L} \nonumber \\
&+\sum_{\{q,q'\}=0}^{L-1} V_{qq'} \hat{b}_{q'}^{\dagger}\hat{b}_{q}
\label{Ham_mom}
\end{align}
where $E_q=-2J\cos(2\pi q/L-\theta)$ are the single-particle energies, $V_{qq'}$ is the Fourier transform of the single-particle potential $V_{qq'} = \frac{1}{L} \sum_j V_j e^{2\pi i(q-q')j/L}$, and the modulus is taken because in collision processes the quasi-momentum is conserved up to an integer multiple of the reciprocal lattice vector $2\pi/d$, i.e.~modulo Umklapp processes. In absence of a potential barrier, i.e.~$V_j=0$, the single-particle spectrum is twofold degenerate for $\theta = \pi/L$ which we will refer to as the critical rotation frequency or critical phase twist.

\begin{figure*}
\centering
\includegraphics[width=0.66\columnwidth]{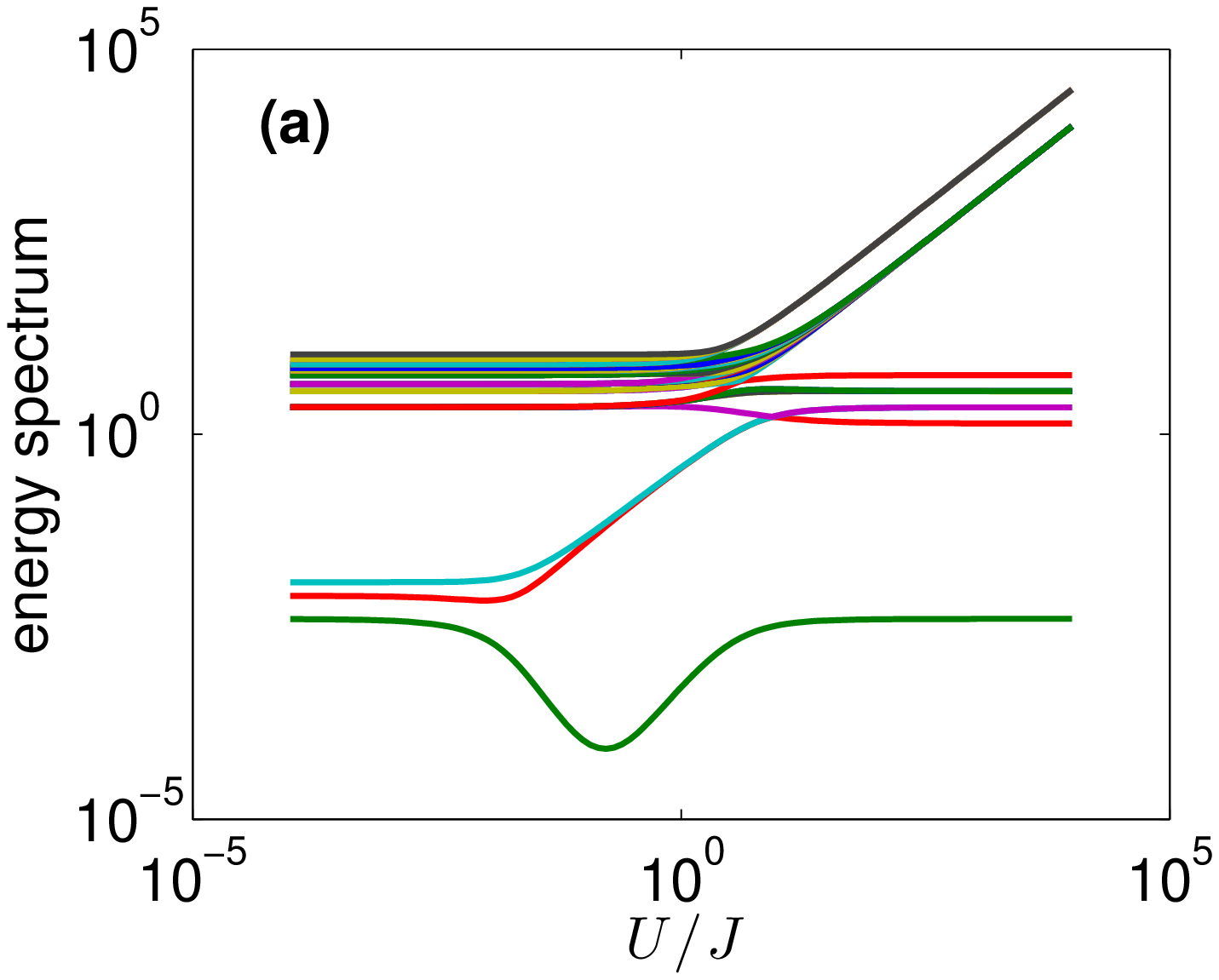}
\includegraphics[width=0.66\columnwidth]{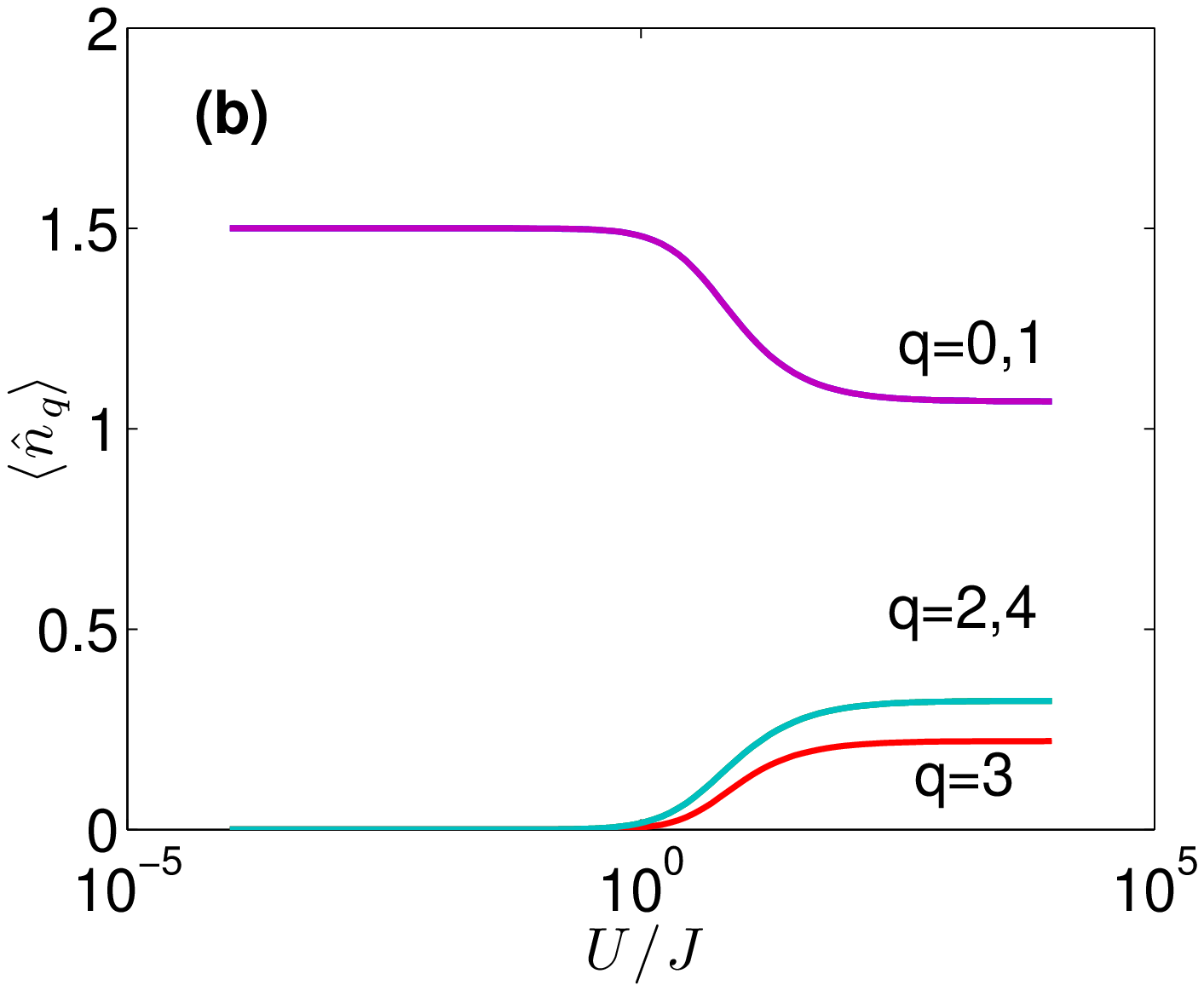}
\includegraphics[width=0.66\columnwidth]{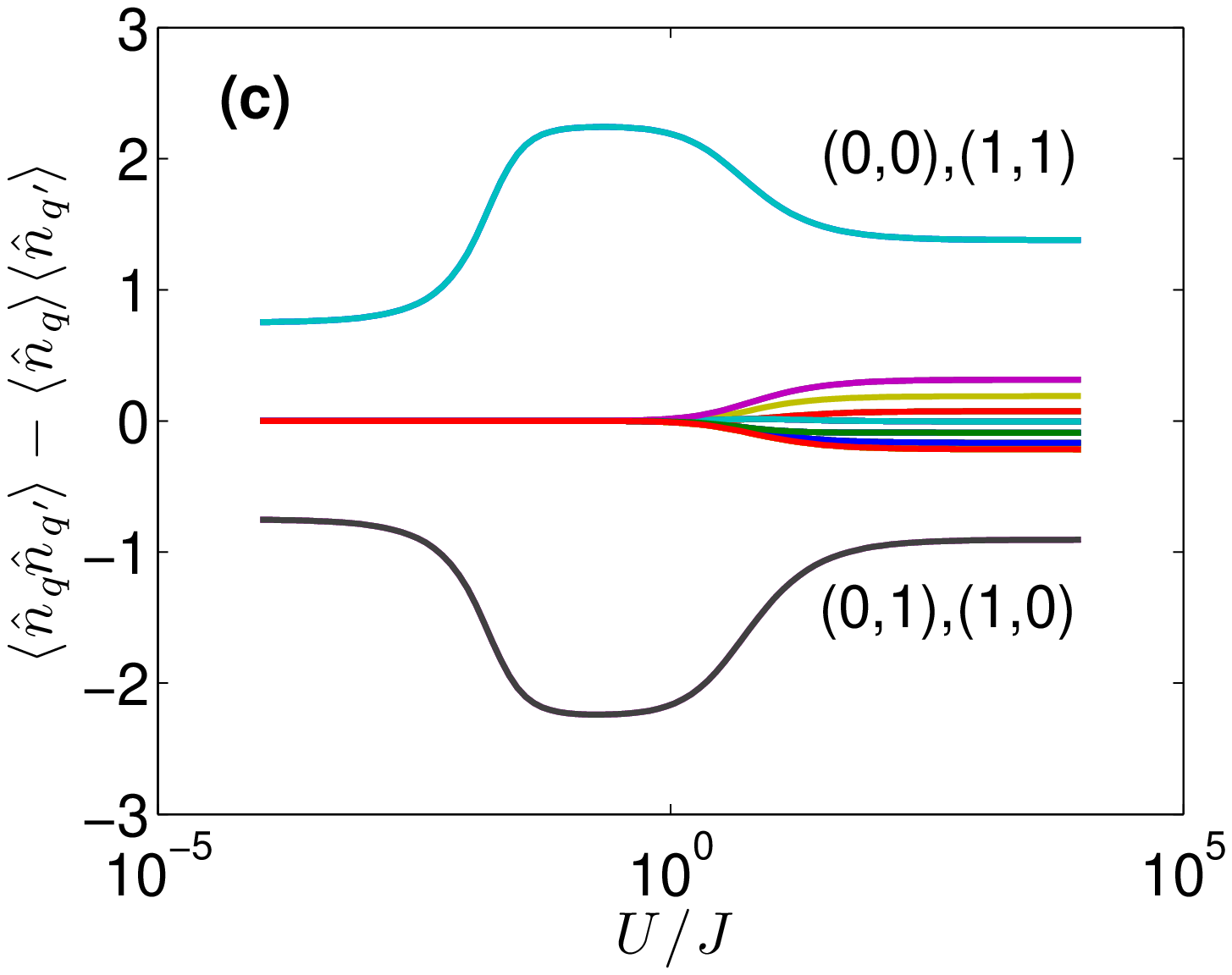}
\includegraphics[width=0.66\columnwidth]{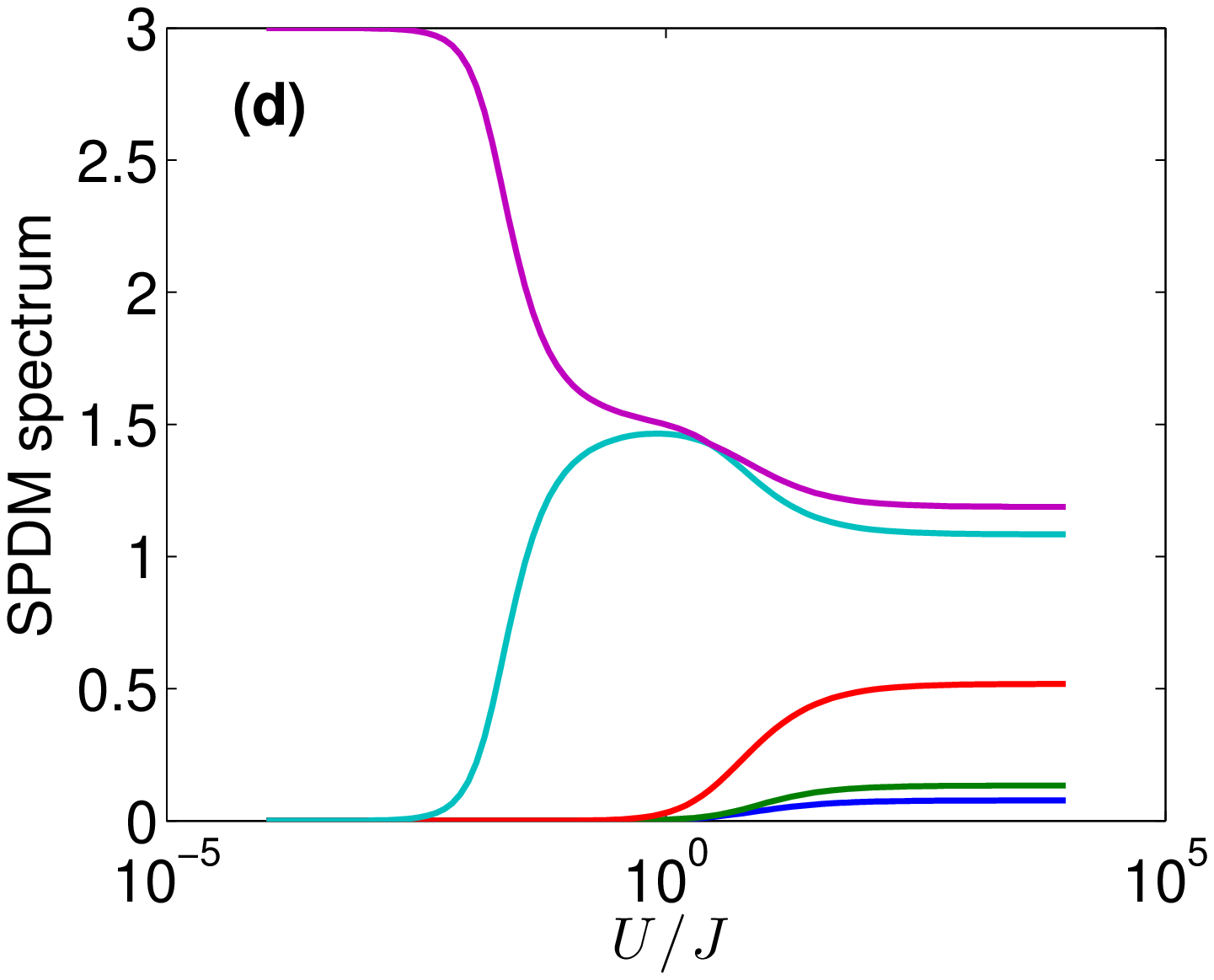}
\includegraphics[width=0.66\columnwidth]{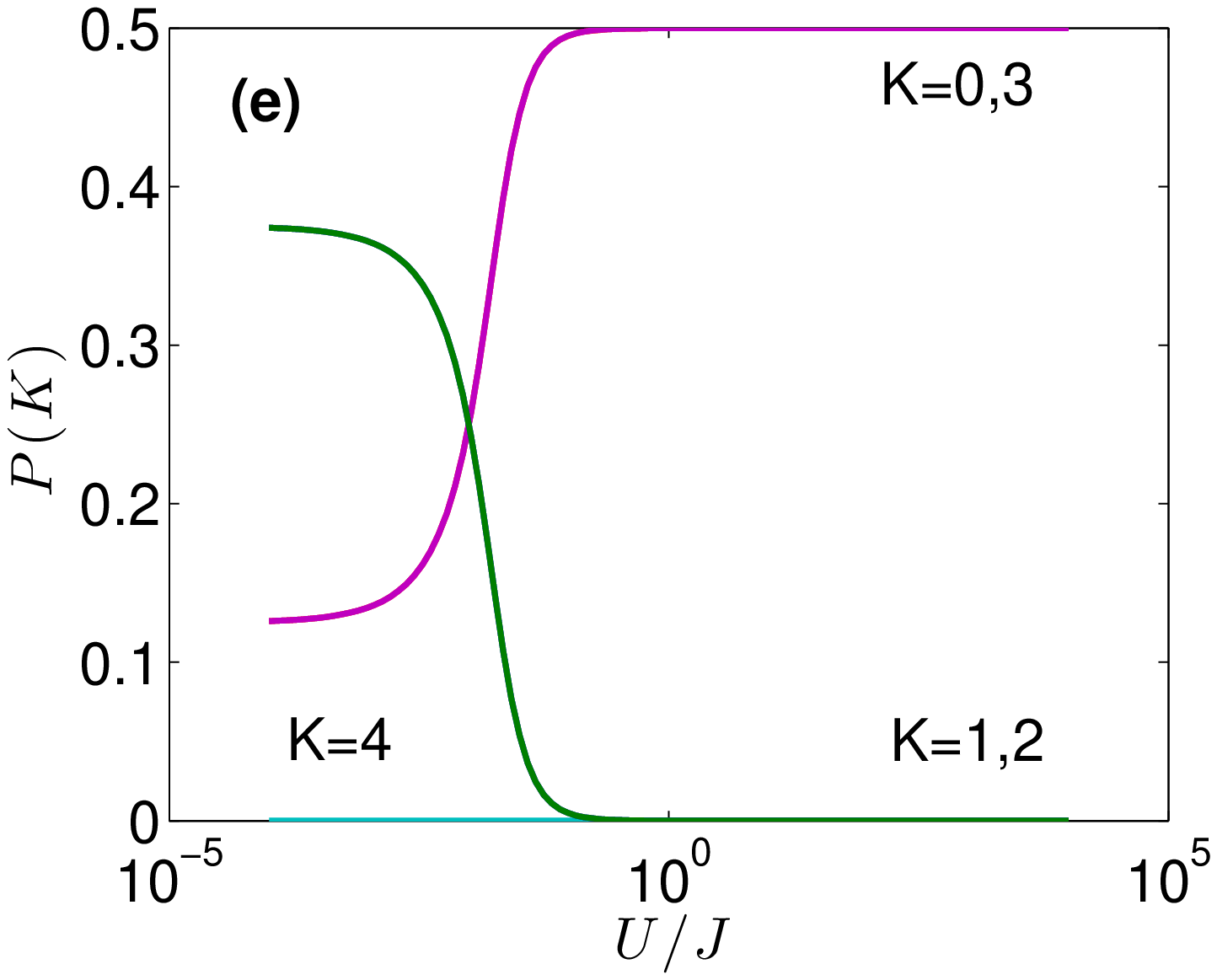}
\includegraphics[width=0.66\columnwidth]{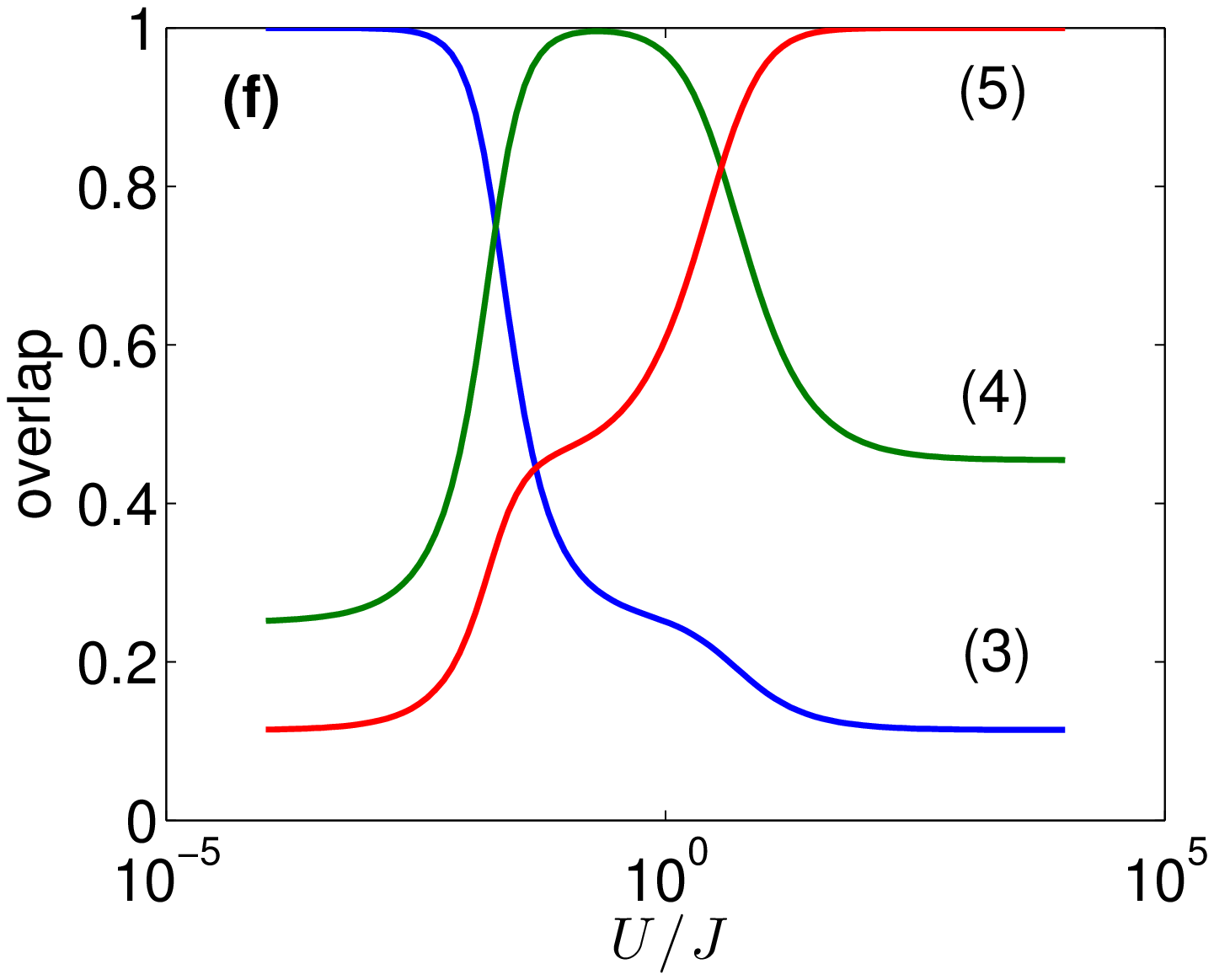}
\caption{(Color online) Energy spectrum and characterization of the ground-state wave function for $N = 3$ particles on $L = 5$ sites at the critical phase twist with $V_0/J = 0.01$ as a function of the interaction parameter $U/J$: (a) many-body energy spectrum, (b) momentum distribution $\langle \hat{n}_q \rangle$, (c) momentum noise correlations $\langle \hat{n}_q \hat{n}_{q'} \rangle - \langle \hat{n}_q \rangle \langle\hat{n}_{q'} \rangle$, (d) spectrum of single-particle density matrix $\langle \cre{a}_i \des{a}_j \rangle$, (e) distribution of total quasi-momentum $P(K)$ (as defined in the text), and (f) fidelity with respect to the states $\ket{\psi_0}$, $\ket{\psi_1}$ and $\ket{\psi_\infty}$ in Eqs.~(\ref{state1}), (\ref{state2}) and (\ref{state3}).}
\label{fig1}
\end{figure*}

\emph{Ground state at the critical phase twist.} In the following we will focus on the ground state at the critical phase twist for a small single-site potential barrier ($V_j = 0$ for $j \not= 0$, i.e.~$V_{qq'}=V_0 \ll J$). Note that a single-site barrier leads to constant momentum transfer in the first Brillouin zone. Similar results were found in Ref.~\cite{Hallwood2010} which we will extend here.

In Fig.~\ref{fig1} we show our results for the many-body spectrum and detailed characterization of the ground-state wave function using exact diagonalization of a small system for $N=3$ atoms on $L=5$ lattice sites. We can clearly distinguish three regimes: weak, intermediate, and strong interactions.

In the non-interacting system, $U = 0$, all bosons occupy the lowest-energy single-particle state. In the presence of the potential barrier, $V_0 \not = 0$, translation symmetry is broken and quasi-momentum ceases to be good quantum number. For a weak barrier, $V_0 \ll J$, the ground state of the single-particle Hamiltonian is an equal superposition of the quasi-momentum states $\ket{q=0}$ and $\ket{q=1}$, and the many-body ground state is
\begin{equation}
| \psi_0 \rangle = \left( \frac{ \cre{b}_0 - \cre{b}_1}{\sqrt{2}} \right)^N \ket{0}
\label{state1}
\end{equation}
where $\ket{0}$ is the many-body vacuum.
This is corroborated by our exact-diagonalization results in the weakly-interacting regime: the momentum distribution is $\langle \hat{n}_q \rangle = \langle \cre{b}_q \des{b}_q \rangle = N/2$ for $q=\{0,1\}$ and $\langle \hat{n}_q \rangle = 0$ otherwise. The distribution of total momentum $P(K)$, i.e.~the probability that the many-body system has total quasi-momentum $2\pi K/dL$ is binomial. It can be calculated from $P(K) = \sum_n \ket{K,n} \bra{K,n}$, where $\ket{K,n} \bra{K,n}$ is the projector on the Fock state with $K$ total quasi-momentum quantum number and $n$ is a shorthand for the remaining quantum numbers~\cite{Hallwood2010}. Finally, the spectrum of the single-particle density-matrix (SPDM) $\langle \cre{a}_i \des{a}_j \rangle$ has one eigenvalue of size $N$ which clearly signals condensation.

As soon as the interactions overcome the energy splitting induced by the potential barrier, the near degenerate many-body states are strongly mixed leading to correlations between the atoms. We discussed this mechanism in detail in the context of NOON-state production in rotating ring lattices and superlattices~\cite{Nunnenkamp2008} as well as quantum vortex nucleation~\cite{Nunnenkamp2010}.
Neglecting the effect of the weak potential barrier, the non-interacting many-body spectrum at the critical phase twist is $N+1$-fold degenerate. Most of the degeneracy is lifted by the interactions in first order of perturbation theory, but the states $(\cre{b}_0)^N \ket{0}$ and $(\cre{b}_1)^N \ket{0}$ remain degenerate. In the presence of the potential barrier these two states are coupled at some higher order of perturbation theory (even if the system is non-commensurate), the degeneracy is lifted, and the ground state is the quasi-momentum NOON-state~\cite{Hallwood2010}
\begin{equation}
| \psi_1 \rangle = \frac{(\cre{b}_0)^N - (\cre{b}_1)^N}{\sqrt{2}} \ket{0}.
\label{state2}
\end{equation}
While the momentum distribution is almost unchanged with respect to the non-interacting case, the momentum noise correlations $\langle \hat{n}_q \hat{n}_{q'} \rangle - \langle \hat{n}_q \rangle \langle\hat{n}_{q'} \rangle$ pick up the large fluctuations in the NOON-state (\ref{state2}). The distribution of total quasi-momentum $P(K)$ is bimodal, and the SPDM has two large eigenvalues of size $N/2$ indicating fragmentation of the condensate. Finally, the overlap with the many-body quasi-momentum basis states unambiguously proofs that the ground state is indeed the quasi-momentum NOON-state (\ref{state2}).

In passing we note that the presence of NOON-states can be observed in time-of-flight expansion after inducing many-body oscillations with a quench in the rotation frequency~\cite{Nunnenkamp08}. Their frequency is given by twice the energy gap which in the regime of intermediate interactions is exponentially small in the number of particles~\citep{Hallwood06a, Rey07}.

Let us now discuss the strongly-interacting regime. In absence of the potential barrier, i.e.~$V_0 = 0$, quasi-momentum is a good quantum number. Strong interactions ($U/J \rightarrow \infty$) lead to fermionization of the bosons within each subspace of total quasi-momentum $K$. In this regime the many-body energy spectrum is equal to the single-particle spectrum (with (anti-)periodic boundary conditions for (even) odd $N$ \cite{Lieb1961}) which is degenerate at the critical phase twist. The potential barrier then breaks translation symmetry and couples the two degenerate ground states $\ket{\psi_\infty^{K=0}}$ and $\ket{\psi_\infty^{K=N}}$, which are the ground states of the quasi-momentum subspaces with $K=0$ and $K=N$. We thus call the ground-state wave function a \emph{center-of-mass superposition}
\begin{equation}
\ket{\psi_\infty} = \frac{\ket{\psi_\infty^{K=0}} - \ket{\psi_\infty^{K=N}}}{\sqrt{2}}.
\label{state3}
\end{equation}
Compared to the regime of intermediate interactions the two largest eigenvalues of the SPDM are smaller in the strongly-interacting limit pointing to the fact that the center-of-mass superposition (\ref{state3}) is only a partially-fragmented state. Since many quasi-momentum basis states contribute to (\ref{state3}) the overlap with a quasi-momentum NOON-state (\ref{state2}) is also smaller.

Please note that the distribution of total quasi-momentum $P(K)$ is not sensitive to the difference between the states (\ref{state2}) and (\ref{state3}). This is why going beyond Ref.~\cite{Hallwood2010} with a more detailed analysis in terms of SPDM and overlap is important.

\begin{figure}
\centering
\includegraphics[width=0.85\columnwidth]{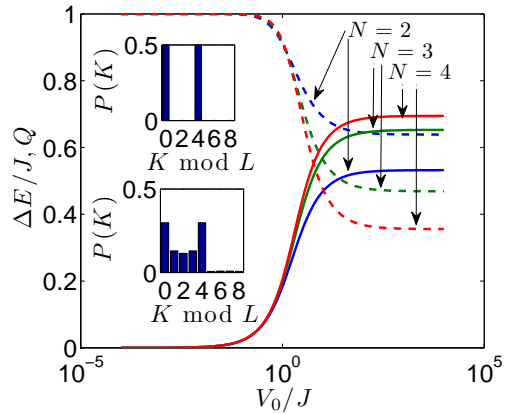}
\caption{(Color online) Energy gap $\Delta E$ (solid) and cat quality $Q = 4 P(0) P(N)$ (dashed) versus single-site barrier height $V_0$ for $L = 9$ as well as $N = 2$ (blue), $N = 3$ (green) and $N = 4$ (red) at the critical phase twist in the strongly-interacting limit $U / J \rightarrow \infty$. Insets show distribution of total quasi-momentum $P(K)$ for $V_0/J=10^{-4}$ and $V_0/J=10^{+4}$ for $N = 4$.}
\label{fig2}
\end{figure}

\emph{Effect of large barrier heights.} So far we have focused on the case of a small single-site potential barrier $V_0 \ll J$. Let us now study how larger barriers affect the system.

In Fig.~\ref{fig2} we plot the energy gap $\Delta E$ and the cat quality $Q = 4 P(0) P(N)$~\cite{Hallwood2010} as a function of the barrier height $V_0$. We find that the gap is small for $V_0 \ll J$, increases sharply around $V_0 = J$, and levels off for $V_0 \gg J$. As Ref.~\cite{Hallwood2010} we see that the energy gap is independent of the number of particles in the small barrier limit and increases with the number of particles in the large barrier limit. However, particularly interesting is the dependence of the cat quality $Q$ and the distribution of total quasi-momentum $P(K)$ as the barrier height increases. Fig.~\ref{fig2} and its insets clearly show that the properties of the superposition states are degraded if the barrier height is raised to increase the energy gap $\Delta E$. This effect becomes more severe as the number of particles increases.

While we study a lattice system which is different from the continuum model considered in Ref.~\cite{Hallwood2010}, we expect that our results qualitatively carry over to the continuum problem as they are obtained at densities well below unit filling.

\begin{figure}
\includegraphics[width=0.8\columnwidth]{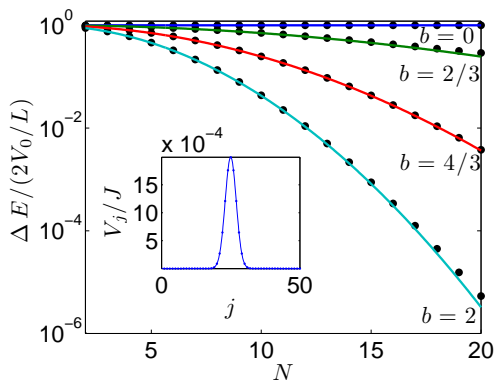}
\caption{(Color online) Energy gap $\Delta E$ versus particle number $N$ for $L = 50$, $V_0/J = 0.01$, $b = \xi/d = 0$ (blue), $b = \xi/d = 2/3$ (green), $b = \xi/d = 4/3$ (red), and $b = \xi/d = 2$ (cyan). Points are the exact single-particle spectrum and lines from perturbative expression (\ref{gap}). The inset shows the barrier potential $V_j$ for $L = 50$, $V_0/J = 0.01$, and $b = \xi/d = 2$.}
\label{fig3}
\end{figure}


\emph{Number scaling for finite-width potential barrier.}
In future experiments the barrier will most likely be provided by a blue-detuned laser beam with Gaussian beam waist. That is why we now generalize the case of a single-site potential barrier and consider a Gaussian potential barrier of finite width $\xi$, i.e.~$V_j = {\cal N}^{-1} V_0 e^{-d^2j^2/2\xi^2}$ with the normalization ${\cal N} = \sum_{j=1}^L e^{-d^2j^2/2\xi^2}$. The matrix elements ${V}_{qq'}$ are given by the discrete Fourier transform of the barrier potential $V_j$ which for $d \ll \xi$ becomes the continuous Fourier transform $V_{qq'} = V_0 e^{-2 \pi^2 (\xi/dL)^2 (q-q')^2}/L$ while for $\xi \rightarrow 0$ we recover the single-site barrier limit $V_{qq'} = V_0/L$.

We can get a perturbative expression for the energy gap using the Bose-Fermi mapping \cite{Lieb1961} and degenerate perturbation theory first order in $V_0$. The energy gap $\Delta E$ between the ground and first excited state in the Tonks limit for $N$ particles is
\begin{equation}
\Delta E = 2 \frac{V_0}{L} e^{-2 \pi^2 (\xi/dL)^2 N^2}.
\label{gap}
\end{equation}
We note that in this weak barrier limit the exponential scaling is independent of the barrier height.

In Fig.~\ref{fig3} we plot the energy gaps $\Delta E$ which open due to the potential barrier as a function of number of particles $N$ for various barrier widths $\xi/d$. In the small barrier $V_0/J = 0.01$ the accuracy of the perturbative expression (\ref{gap}) is excellent. In contrast to the single-site potential barrier, i.e.~$\xi/d \ll 1/L$, a potential barrier with finite width $\xi$ leads to exponentially small energy gaps for increasing number of particles $N$. The physical reason is that we need increasingly large momentum transfer to couple the center-of-mass momentum states. Since the matrix elements of a finite-width potential barrier decrease exponentially for increasing momentum transfer, we recover the exponential scaling of the energy gap $\Delta E$ similar of the weakly-interacting regime.

The NIST group \cite{Ramanathan2011} recently carried out experiments in a toroidal trap with radius of $20 \mu\textrm{m}$ and a $4.3 \mu\textrm{m}$ ($1/e^2$ radius) repulsive barrier. This corresponds to $\xi/dL \approx 1/60$ and a suppression of the gap (\ref{gap}) by $e^{-(N/10)^2/2}$. This will severely limit cat state production beyond few tens of particles in the weak barrier regime where the cat quality is good.

\emph{Summary.} In conclusion, we have shown that a weak potential barrier in a rotating ring lattice loaded with ultracold bosons at low filling fraction supports several qualitatively very different superposition states of different total quasi-momentum: a condensate for weak, a NOON-state for intermediate, and a center-of-mass superposition for strong interactions. Raising the potential barrier will increase the energy gap but degrade the superposition states. The energy gap in the strongly-interacting limit is only independent of the number of particles as long as one can neglect the width of the potential barrier. Otherwise one recovers the exponential scaling with increasing number of particles. These are crucial issues in the production of superposition states in rotating atomic rings. They severely limit the size of superposition states in any realizable experimental system.

\emph{Acknowledgements.} AN thanks Steven M.~Girvin for a careful reading of the manuscript and acknowledges support from NSF under grant DMR-1004406. AMR acknowledges support from NSF, NIST and a grant from the ARO with funding from the DARPA-OLE.

\end{document}